\documentclass[aps,twocolumn,showpacs,superscriptaddress,preprintnumbers,floatfix,amsmath,amssymb,eufrak,table]{revtex4-1}
\usepackage{ulem}
\usepackage{color}
\usepackage{graphicx}  
\usepackage{appendix}
 \usepackage{epsfig}
\usepackage[colorlinks=true,linkcolor= red, citecolor = blue, urlcolor = teal ]{hyperref}
\usepackage{epstopdf}
\usepackage{amsmath}
\usepackage{subfig}
\usepackage{comment}
\usepackage{float}
\usepackage[dvipsnames]{xcolor}
\def\bea {\begin{eqnarray}}
\def\eea {\end{eqnarray}}

\newcommand{\be}{\begin{equation}}
\newcommand{\ee}{\end{equation}}
\newcommand{\ba}{\begin{eqnarray}}
\newcommand{\ea}{\end{eqnarray}}
\newcommand{\nn}{\nonumber\\}

 \def\om{\omega}

\DeclareGraphicsExtensions{.jpg,.pdf,.eps}

\begin{document}

\title{Impact of chiral asymmetry and magnetic field on passage of an energetic test parton in a QCD medium}

\author{Ritesh Ghosh}
	\email{riteshghosh1994@gmail.com}
	\affiliation{School of Physical Sciences, National Institute of Science Education and Research, An OCC of Homi Bhabha National Institute,  Jatni, Khurda 752050, India}

\author{Mohammad Yousuf Jamal}
\email{mohammad@iitgoa.ac.in}
\affiliation{School of Physical Sciences, Indian Institute of Technology Goa, Ponda 403401, Goa, India}

\author{Manu Kurian}
\email{mkurian@bnl.gov}
\affiliation{RIKEN BNL Research Center, Brookhaven National Laboratory, Upton, New York 11973, USA}

\begin{abstract}
We study the dependence of collisional energy loss of a test parton moving with a high velocity on the chiral imbalance and magnetic field in the QCD medium. A semi-classical approach is adopted to estimate the parton energy loss that takes into account the back-reaction on the parton due to the polarization effects of the QCD medium while traversing through the medium. We find that the motion of the parton is sensitive to the chiral asymmetry in the medium. Further, we investigate the effect of magnetic field-induced anisotropy on the energy transfer between the moving parton and the medium. Our results show that the energy loss of the parton is strongly influenced by the strength of the magnetic field as well as the relative orientation of the motion of the parton and the direction of the magnetic field in the medium.\\
\end{abstract}
	\maketitle 
	\newpage

\section{Introduction} 
The heavy-ion collision (HIC) experiments at Relativistic Heavy Ion Collider (RHIC) and Large Hadron Collider (LHC) provide a unique opportunity to produce and investigate the properties of an extremely hot and dense state of matter: the Quark-Gluon Plasma (QGP), which is believed to resemble the state of the universe shortly after the Big Bang~\cite{PHOBOS:2004zne, BRAHMS:2004adc, ALICE:2010khr, Fukushima:2020yzx}. The success of relativistic viscous hydrodynamics in describing the evolution of the QGP medium is considered one of the theoretical breakthroughs in this area of research~\cite{Gale:2013da, Heinz:2013th}. The dissipative processes and the associated transport parameters rely on the understanding of the non-equilibrium physics of the QGP. Much research has been devoted to the extraction of the transport coefficients by employing the hydrodynamics approach and data fitting methods to the observables associated with the final state particles~\cite{Romatschke:2007mq, Ryu:2015vwa}. The data-driven methods are currently getting wider attention in this field as the physics of the QGP is entering a high-precision science era~\cite{Bernhard:2019bmu, JETSCAPE:2020shq, Heffernan:2023utr}.

The class of {\it hard probes} such as jets, heavy quarks, {\it etc} offers another possible way to characterize the properties of the QGP. These energetic patrons penetrate and travel through the QGP and lose their energy due to several interactions in the medium. Therefore, proper knowledge of the energy loss of a fast-moving parton is essential for the quantitative description of the jet quenching phenomenon. The theoretical modeling of the energy loss mechanism of the energetic partons and its dependence on various observables of collision experiments play a significant role in elucidating the underlying physics. Several efforts have been made to understand the transport properties of the QGP through the investigation of jets while considering both the collisional and radiative processes of the energetic parton~\cite{Mustafa:2004dr, Sarkar:2018erq, Liu:2020dlt, Mazumder:2013oaa, Prakash:2021lwt, Shaikh:2021lka, Dai:2020rlu}. The heavy quarks are another promising candidates that can probe the evolution history of the QGP as they are created in the very initial stages of the HIC~\cite{vanHees:2005wb, Das:2009vy, Das:2013kea, Cao:2018ews, Song:2020tfm, Kurian:2020orp, Sebastian:2022sga}. The motion of the energetic partons and their energy loss depend on the properties of the medium. There have been some attempts to explore the  parton interaction with the non-equilibrium plasma or unstable medium~\cite{Dumitru:2007rp, Carrington:2015xca, Jamal:2020fxo, Hauksson:2021okc, Carrington:2022bnv, Ruggieri:2022kxv, Hauksson:2023tze}. 

The chiral anomaly and parity-violating effects have recently attracted substantial theoretical and experimental interest in the study of the QGP and strong interaction. The interaction of the chiral fermions/quarks with the nontrivial gluonic field produces asymmetry between left- and right-handed chiral fermions that yield chiral imbalance~ \cite{KHARZEEV20161}. Furthermore, the recent observations~\cite{Acharya:2019ijj, Adam:2019wnk} and the associated studies~\cite{Das:2016cwd, Jiang:2022uoe} of the enhanced directed flow of heavy flavor mesons revealed the existence of a strong magnetic field (for reviews on the strong magnetic field in collision experiments, see Refs.~\cite{Miransky:2015ava, Huang:2015oca, Hattori:2016emy, Kharzeev:2015znc}) in the initial stages of the HICs. This generated magnetic field breaks the rotational symmetry of the medium, and induces anisotropy in the produced medium that, in turn, affects the thermodynamic and transport properties of the QCD medium as the charged particle motion is influenced by the field~\cite{Karmakar:2019tdp, Hattori:2022hyo, Das:2021fma, Dash:2020vxk,Wang:2022jxx, Kurian:2018qwb, Ghosh:2018cxb, Gowthama:2020ghl, Rath:2020idp, Wang:2020dsr, Das:2019ppb, Ghosh:2022vjp}. Refs.~\cite{Fukushima:2015wck, Singh:2020faa,Mazumder:2022jjo} showed that the charm quark momentum diffusion critically depends on the preferred direction and strength of the magnetic field. The magnetic field, together with the chiral imbalance, leads to novel phenomena such as Chiral Magnetic Effect (CME) and is one of the active areas of ongoing studies in contemporary physics~\cite{Fukushima:2008xe, Kharzeev:2010gr, Sadofyev:2010pr, Kharzeev:2015znc, An:2021wof}. Both the magnetic field and chiral imbalance are expected to influence the passage of partons through the medium, as the momentum broadening and energy loss of energetic partons depend on the properties of the medium. It is an exciting task to study the impact of chiral asymmetry and magnetic field on the energy loss pattern of the moving parton in the medium, and this sets the motivation for the present study.

A semi-classical approach is utilized to set up the formalism for energy loss experienced by an energetic parton that incorporates its interactions with the chromodynamic fields in the QCD medium. The motion of the test parton in plasma is described with Wong's equations by treating it as a classical particle with ${SU(N_c)}$ color charge. The change in the color field configuration due to the passage of the parton is embedded through the linearized Yang-Mills equations. We consider three different choices of plasma, namely {\it (i)} isotropic, {\it (ii)} chiral asymmetric, and {\it (iii)} magnetized QCD medium. While traversing the medium, the back-reaction exerted on the parton medium is taken into account by analyzing the polarization effects of the medium. We observe that the chiral asymmetry affects the parton energy loss mechanism. Further, we show that the presence of the magnetic field suppresses the energy loss. Another crucial finding is the magnetic field-induced anisotropy in the parton energy loss, which indicates that the relative orientation of the parton's motion and the magnetic field's direction strongly influences the energy loss.   

We organize the manuscript as follows. In section~\eqref{sec:EL}, we present the formalism for the energy loss of an energetic parton moving in three different choices of QCD medium. Section~\eqref{Res} is devoted to the results and discussions. Finally, we summarize the analysis with an outlook in section~\eqref{Summ}.

{\it{ Notations and conventions}:}  In the present analysis, we consider $c=k_B=\hbar=1$, $g_{\mu\nu}={\text {diag}}\,(1,-1,-1,-1)$, $N_c=3$, and $N_f=2$. The quantity $q_{f}$ denotes the particle's electric charge with flavor $f$. A four-vector is defined as $X^{\mu}=(x^0, |\bf{x}|)$ with the component of the three-vector is described with the Latin indices $x^i$ where $i=(1,2,3)$.

\section{Formalism}\label{sec:EL}
{A highly energetic test parton, while passing through the QGP medium, loses energy due to its interactions with the color fields. The energy loss experienced by the parton can be measured through the work of the retarding forces acting on the parton in
the medium from the induced chromo-electric field that generated due to its motion.
The dynamics of the test parton in chromodynamic fields can be described with Wong's equations~\cite{Wong:1970fu} that in the Lorentz covariant form given as,    
    \begin{align}
    \frac{dX^{\mu}(\tau)}{d\tau} &= V^{\mu}(\tau),\label{eq:1_1} \\
    \frac{dP^{\mu}(\tau)}{d\tau} &=g_s q^{a}(\tau)F^{\mu\nu}_{a}(X(\tau)){V}_{\nu}(\tau),\label{eq:1_2} \\
    \frac{dq^{a}(\tau)}{d\tau} &= -g_sf^{abc}V_{\mu}(\tau)A^{\mu}_{b}(X(\tau))q_{c}(\tau),
    \label{eq:wong}
    \end{align} 
with $\tau$, $X^\mu (\tau)$,  $V^\mu (\tau)$ and  $P^\mu (\tau)$ as the proper time, position, velocity, and momentum of the test parton. Here, $g_s$ denotes the strong running coupling constant; $q_a$ is the parton color charge; $F^{\mu\nu}$ represents chromodynamic field tensor; $A^\mu$ is the gauge potential; $f^{abc}$ describes the structure constant  of ${SU}( N_c)$ group; and $a$ defines the color index with $a=1,2,\dots, N_c^2-1$. We follow two assumptions to solve Wong's equations; first, we choose the gauge condition $V_{\mu}A^{\mu}_{a}(X) = 0$, which indicates the gauge potential vanishes on the particle's trajectory~\cite{Jiang:2016duz, Carrington:2015xca}. Secondly, we consider that the energy of the moving parton is comparatively much larger than its energy loss in the medium~\cite{Jiang:2016duz, Chakraborty:2006db}. Next, considering $\mu=0$ component in Eq.~\eqref{eq:1_2} and replacing proper time with time $t=\gamma\tau$, we get,
    \ba
    \dfrac{d{\it E}}{dt}&=&g_{s}q^a {{\bf v}}\cdot { {\textbf E}_{a}}(t, {\bf x}={\bf v}t),
    \label{eq:el1st}
    \ea 
where ${\textbf E}_{a}(t,{\bf x}={\bf v}t)$ is the chromo-electric field induced due to the motion of the parton with the energy, $E$  and velocity, ${\bf v}$. The energy loss can also be described in terms of the color current ${\bf j}_a$ generated by the energetic moving parton as,
    \ba
    \frac{d{\it E}}{dt}&=&\int d^3{\bf x}\, { {\textbf E}_{a}}(t, {\bf x})\cdot {\bf j}_a (t, {\bf x}),
    \label{eq:el1}
    \ea 
with ${\bf j}_a (t, {\bf x})=g_{s}q^a {{\bf v}}\delta^{(3)}({\bf x}-{\bf v}t)$. The form of ${\textbf E}_{a}$ generated due to the test parton motion in the medium can be obtained by solving the linearized Yang-Mills equation. Employing the conventional way, {\it i.e.,} Fourier transforming the linearized differential equation to algebraic forms, the induced chromo-electric can be obtained as~\cite{Thoma:1990fm, Chakraborty:2006md},
\begin{align}
{\text E}^{i}_a(K)=i\omega \Delta^{ij}(K)j^{j}_a(K),
\label{eq:eind}
\end{align}
where {$K^\mu=({\omega,{\bf k}})$.} The gluon propagator $\Delta^{ij}(K)$ and the external current $j^{j}_a(K)$ take the following forms~\cite{YousufJamal:2019pen},
\begin{align}
&\Delta^{ij}= [(|{\bf k}|^2-\omega^2)\delta^{ij}-k^ik^j+\Pi^{ij}(K)]^{-1}\label{eq:jind10},\\
&j^{j}_a(K)=\frac{i g_s q^a v^j}{\omega -{\bf k}\cdot{\bf v}+i 0^+}.
\label{eq:jind1}
\end{align}
Here, $\Pi^{ij}(K)$ is the gluon self-energy tensor that captures the medium effects. Using Eq.~\eqref{eq:jind10} and Eq.~\eqref{eq:jind1} in Eq.~\eqref{eq:eind}, one can obtain the induced field, and that, in turn, gives the change of parton energy in the Fourier space. Next, to get back to the coordinate space, we perform the inverse Fourier transformation, which after averaging over color indices and completing $\omega-$integration using the Residue theorem (as the integrand has a pole at $\omega ={\bf k}\cdot{\bf v}$)  can be written as,
    \begin{align}
    \bigg\langle{\frac{d{ E}}{dx}}\bigg\rangle=i\frac{1}{|{\bf v}|}g_s^2C_F v^i v^j \int{\frac{d^3{\bf k}}{(2\pi)^3}\omega \Delta^{ij}},
    \label{main}
    \end{align} 
where $C_F$ is the Casimir invariant of $SU(N_c)$ and ${\omega = {\bf k}\cdot{\bf v}}$. Notably, the change of energy of the parton depends on its velocity and the gluon propagator. The latter has a strong dependence on the choice of the medium. Depending on the medium properties and initial conditions, the energy change can be negative or positive. If parton loses energy while interacting with the medium, $-\big\langle{\frac{d{ E}}{dx}}\big\rangle$ should be positive; otherwise, if in some special cases, the test parton gains energy, $-\big\langle{\frac{d{ E}}{dx}}\big\rangle$ should be negative~\cite{Chakraborty:2006db, Jamal:2020fxo}. Next, we shall discuss the gluon propagator (that depends of the gluon self-energy or the dielectric permittivity of the medium) and the energy loss mechanism for three different systems.

\subsubsection{Moving parton in isotropic medium}
 For an isotropic medium, the gluon self-energy tensor can be decomposed in terms of longitudinal projection operator $B^{\text{ij}}=\frac{k^i k^j}{|{\bf k}|^2}$ and transverse projection operator $A^{\text{ij}}=\delta ^{\text{ij}}-\frac{k^i k^j}{|{\bf k}|^2}$ as,
\ba
\Pi^{ij}(K)= A^{ij}~\Pi_T(K)+B^{ij}~\Pi_L(K),
    \label{eq:eplt1}
\ea  
where $\Pi_L$ and $\Pi_T$ are the longitudinal and transverse form factors, respectively. The $\Pi^{ij}(K)$ can be related to the induced current in the medium via linear response theory. The current induced in the medium can be quantified in terms of the deviation of the medium particle distribution functions that can be obtained by solving the Boltzmann-Vlasov transport equation. Hence, the form factors can be obtained by solving the transport equation and employing linear response theory, and can be expressed as~\cite{Bellac:2011kqa, Mustafa:2022got,Carrington:2021bnk},
\begin{align}
    &\Pi _T(K)={m_D^2}\frac{\omega^2}{2|{\bf k}|^2} \bigg[1-\frac{K^2}{2\omega |{\bf k}|}\ln \left(\frac{\omega+|{\bf k}| }{\omega-|{\bf k}|}\right)\bigg],
    \label{eq:eL}\\
    &\Pi _L(K)=-m_D^2\frac{\omega^2}{k^2}\left[1-\frac{\omega}{2k} \ln \left(\frac{\omega+k }{\omega-k}\right)\right],
    \label{eq:eT}
    \end{align}	
where $K^2=\omega^2-|{\bf k}|^2$ and $m_D^2=(N_f+2N_c)\frac{g_s^2T^2}{6}$ is the Debye screening mass. Employing Eq.~\eqref{eq:eplt1} in Eq.~\eqref{eq:jind10}, the gluon propagator for an isotropic medium can be written as,
\begin{align}
\Delta^{ij}=\frac{1}{C_T}~A^{ij}+ \frac{1}{C_L}~B^{ij}\label{eq:jind101},
\end{align}
with $C_T=-\omega^2+|{\bf k}|^2+\Pi_T$ and $C_L=-\omega^2+\Pi_L$. By substituting Eq.~\eqref{eq:jind101} in Eq.~\eqref{main}, the loss of the moving parton in an isotropic medium can be obtained as, 
    \begin{align}
    -\bigg\langle{\frac{d{ E}}{dx}}&\bigg\rangle=\frac{ \alpha_s C_F}{2 \pi ^2 |{\bf v}|}
    \int d^3{\bf k}\frac{\omega }{|{\bf k}|^2}\bigg\{\omega^2\text{Im}\Big(-\omega ^2+\Pi_L\Big)^{-1}\nonumber\\ 
    &+\left(|{\bf k}|^2 |{\bf v}|^2-\omega ^2\right)\text{Im}\Big({-\omega ^2+k^2+\Pi_T}\Big)^{-1}\bigg\}_{\omega = {\bf k}\cdot{\bf v}}.
    \label{eq:de}
    \end{align} 
The medium effects are incorporated through the form factors of the gluon self-energy and coupling constant. We can rewrite Eq.\eqref{eq:de} (as estimated in other parallel studies \cite{Thoma:1990fm, Jamal:2020emj, Prakash:2023hfj}) in terms of the longitudinal $\epsilon_L(K)$ and transverse $\epsilon_T(K)$ components of the dielectric permittivity $\epsilon^{ij}(K)$ of the medium as,
\begin{align}
     -\bigg\langle{\frac{d{ E}}{dx}}\bigg\rangle &=\frac{\alpha_s C_F}{2 \pi ^2 |{\bf v}|}
    \int d^3{\bf k}\frac{\omega }{|{\bf k}|^2}\bigg\{\text{Im}\Big({\epsilon_{L}(K)}\Big)^{-1}\nonumber\\ 
    &+\left(|{\bf k}|^2 { |{\bf v}|}^2-\omega ^2\right)\text{Im}\Big({\omega ^2 \epsilon_T(K)-|{\bf k}|^2}\Big)^{-1}\bigg\}_{\omega = {\bf k}\cdot{\bf v}},\nn
    \label{eq:de1}
\end{align} 
where the relation between gluon self-energy and dielectric permittivity is taken as,
 \ba
 \epsilon^{ij}(K)=\delta^{ij}-\frac{1}{\omega^2}\Pi ^{\text{ij}}(K).    
 \ea
   It is important to emphasize that the Eq.~(\ref{eq:de}) and Eq.~(\ref{eq:de1}) describe the energy loss due to the polarization effects of the medium, $i.e$, the change in parton energy due to its interaction with collective excitations of the medium. For the numerical calculation, we took the upper limit of the integration $k_{\text{max}}\sim E$, $i.e.$, the initial energy of the parton.
{\subsubsection{Moving parton in chiral imbalance medium}}
The asymmetry between right-handed and left-handed fermions can be quantified in terms of the chiral chemical potential, $\mu_5\equiv \mu_R-\mu_L$. The chiral plasma with a finite $\mu_5$ can be described within the kinetic theory with Berry curvature terms~\cite{Stephanov:2012ki, Son:2012wh, Chen:2016xtg, Akamatsu:2013pjd}. In the small gauge field $A^\mu$, using the linear response theory, we can express $\Pi^{\mu\nu}$ in terms of induced current density in the chiral medium, which can be obtained from the linearized transport equation with Berry curvature correction. { The parity-violating terms are entering through the Berry curvature terms, and $\Pi^{ij}$ with finite $\mu_5$ can be described as~\cite{Akamatsu:2013pjd},
\begin{align}
    \Pi^{ij}(K)= \Pi^{ij}_+(K)+ \Pi^{ij}_-(K),
\end{align}
where $\Pi^{ij}_+(K)$ and $\Pi^{ij}_-(K)$ are parity even and parity odd parts of self-energy respectively, and can be defined as,
\begin{align}
    &\Pi^{ij}_+(K)=\frac{m_D^2\omega}{4\pi}\int d{\bf u} \frac{u^i u^j}{U\cdot K+i\epsilon},\\
    & \Pi^{ij}_-(K)=\frac{\mu_5g_s^2}{4\pi^2}i\epsilon^{ijk}k^k\bigg(1-\frac{\omega^2}{|{\bf k}|^2}\bigg)\bigg[ 1-\frac{\omega}{2|{\bf k}|} \ln \left(\frac{\omega+|{\bf k}| }{\omega-|{\bf k}|}\right)\bigg].
\end{align}
Here, $U^\mu=(1, \bf{u})$ with $\bf{u}=\frac{\bf{k}}{|{\bf k}|}$, the indices $i,j,k$ represent the spatial components}, and the Debye mass in the chiral medium takes the form as, 
\begin{align}\label{Ch6}
m_D^2=(N_f+2N_c)\frac{g_s^2T^2}{6}+N_f\frac{g_s^2}{2\pi^2}(\mu_R^2+\mu_L^2).
\end{align}
Due to the parity-violating term, an additional projector operator is required for the tensorial decomposition of $\Pi^{ij}$ in the chiral medium as compared to the isotropic case. At finite $\mu_5$, gluon propagator can be written as~\cite{Akamatsu:2013pjd},
\begin{align}
\Delta^{ij}=\frac{C_T}{C_T^2-C_A^2}~A^{ij}+ \frac{1}{C_L}~B^{ij}- \frac{C_A}{C_T^2-C_A^2}~C^{ij}\label{Ch2},
\end{align}
where $C^{ij}=i\epsilon^{ijk}\frac{k^k}{|{\bf k}|}$ is the anti-symmetric tensor and $C_A$ that denotes the partity violating component of the self-energy  takes the form as,
\begin{align}\label{Ch5}
C_A\equiv \Pi_A=\mu_5\frac{g_s^2|{\bf k}|}{4\pi^2}\bigg(1-\frac{\omega^2}{|{\bf k}|^2}\bigg)\bigg[ 1-\frac{\omega}{2|{\bf k}|} \ln \left(\frac{\omega+|{\bf k}| }{\omega-|{\bf k}|}\right)\bigg].
\end{align}
Notably, $C_A$ is proportional to chiral chemical potential and in the limit $\mu_5=0$, Eq.~(\ref{Ch2}) reduce back to the result of isotropic case as described in Eq.~(\ref{eq:jind101}). Energy loss of the parton can be influenced by the chiral asymmetry in the medium and can be quantified in terms $\Delta^{ij}$. Employing Eq.~(\ref{Ch2}) in Eq.~(\ref{main}), the energy loss of the parton in a chiral imbalance plasma can be written as,
\begin{align}
    \bigg\langle{\frac{d{ E}}{dx}}\bigg\rangle=&\frac{ig^2_sC_F}{{ |{\bf v}|}}
    \int \frac{d^3{\bf k}}{(2\pi)^3}\frac{\omega }{|{\bf k}|^2}\bigg\{\omega^2\frac{1}{C_L}+\Big(|{\bf k}|^2{ |{\bf v}|}^2-\omega ^2\Big)\nonumber\\
    &\hspace{-.8 cm}\times\frac{C_T}{C_T^2-C_A^2}-i |{\bf k}|~{\bf v}\cdot ({\bf v}\times {\bf k})\frac{C_A}{C_T^2-C_A^2}\bigg\}_{\omega = {\bf k}\cdot{\bf v}}.
    \label{Ch3}
\end{align} 
By substituting the forms of $C_L$, $C_T$, and $C_A$ in Eq.~(\ref{Ch3}) we obtain,
\begin{align}
    -\bigg\langle{\frac{d{ E}}{dx}}\bigg\rangle=&\frac{ \alpha_s C_F}{2 \pi ^2 { |{\bf v}|}}
    \int d^3{\bf k}\frac{\omega }{|{\bf k}|^2}\bigg\{\omega^2\text{Im}\Big(-\omega ^2+\Pi_L\Big)^{-1}\nonumber\\
    &\hspace{-1.2 cm}+\left(|{\bf k}|^2 { |{\bf v}|}^2-\omega ^2\right)
    \text{Im}\Big(\frac{-\omega ^2+|{\bf k}|^2+\Pi_T}{(-\omega ^2+|{\bf k}|^2+\Pi_T)^2-\Pi_A^2}\Big)\bigg\}_{\omega = {\bf k}\cdot{\bf v}}
    \label{Ch4}
 \end{align} 
 The energy loss term that originates from contraction of the anti-symmetric projection operator $C^{ij}$ given in Eq.~\eqref{Ch2} vanishes, however, the chiral effects still enter through the Debye mass and partity violating scalar component of the self-energy $\Pi_A$. In the case of vanishing chiral chemical potential, Eq.~(\ref{Ch4}) reduces back to the form of test parton energy loss expression for an isotropic case as defined in Eq.~\eqref{eq:de}.
 
{\subsubsection{Moving partons in magnetized QCD medium}}
	\begin{figure}
		\begin{center}
			\includegraphics[scale=0.52]{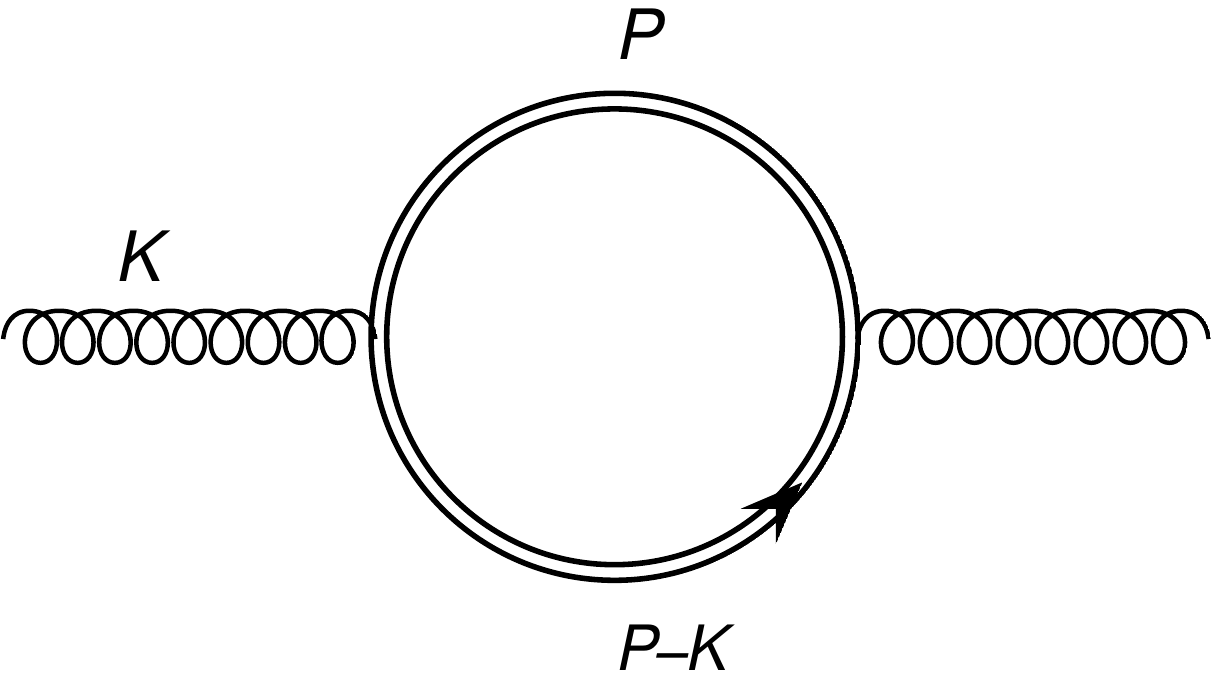}
			\caption{ One loop gluon self-energy. The double
line denotes the modified quark propagator in the presence of a strong magnetic field}
			\label{fig:se_B0}
		\end{center}
	\end{figure}
The magnetic field is a source of anisotropy in the system, which will be reflected in the medium's QCD thermodynamics and transport processes. The motion of the test parton will be affected by the strength ${\bf B}$ and direction of the magnetic field ${\bf n}$ in the medium. In the magnetized medium, the tensor basis for a symmetric second-rank tensor will be different from that of an isotropic medium due to the inclusion of the additional magnetic field vector. Using vectors $k^{i}$ and $n^i$ with $n^2=1$, we can construct the spacial part of the gluon self-energy. In the analysis, we consider various cases with different directions of the test parton velocity ${\bf v}$ to analyze the impact of the relative orientation of ${\bf v}$ and ${\bf n}$ on the parton energy loss. In the presence of a magnetic field, $\Pi^{ij}$ can be decomposed in terms of basis tensors which depend on $k^i$ and $n^{i}$ as,
	\bea
	\Pi^{ij}=a\,N^{ij}+ b \,B^{ij}+c \,R^{ij}+d \,Q^{ij},\label{self_energy}
	\eea
	where the projection operators can be defined as follows,
	\begin{align}
	&N^{ij}=-\frac{\hat k^i \tilde n^j+\hat k^j \tilde n^i}{\sqrt{\tilde n^2}},\\
   &B^{ij}=\frac{k^i k^j}{|{\bf k}|^2},\\
	&R^{ij}=-\delta^{ij}+\frac{k^i k^j}{|{\bf k}|^2}-\frac{\tilde n^i \tilde n^j}{\tilde n^2}
	,\\
&Q^{ij}=\frac{\tilde n^i \tilde n^j}{\tilde n^2}.  
	\end{align}
\begin{figure*}
		\begin{center}
			\includegraphics[scale=0.55]{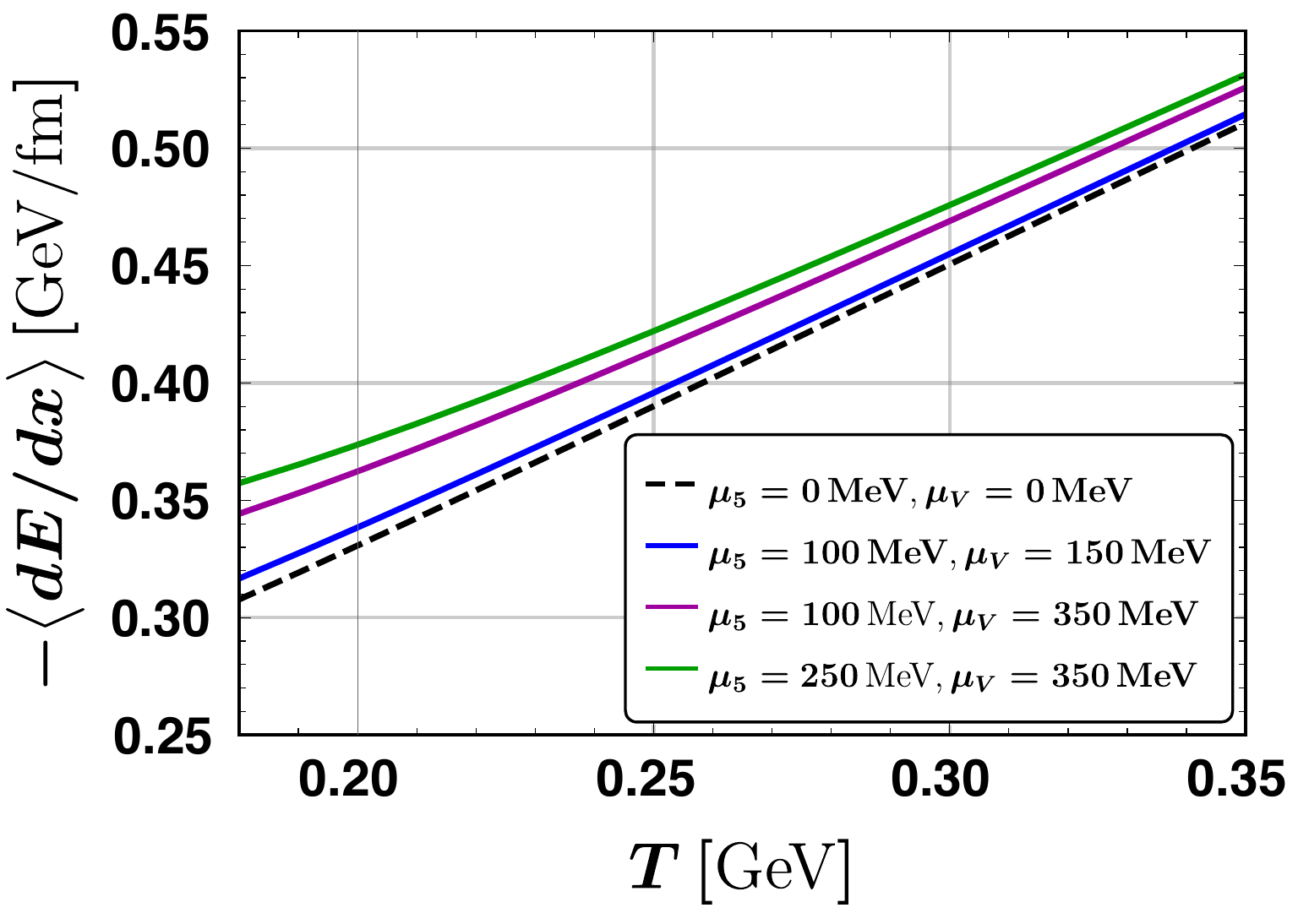}
   \includegraphics[scale=0.58]{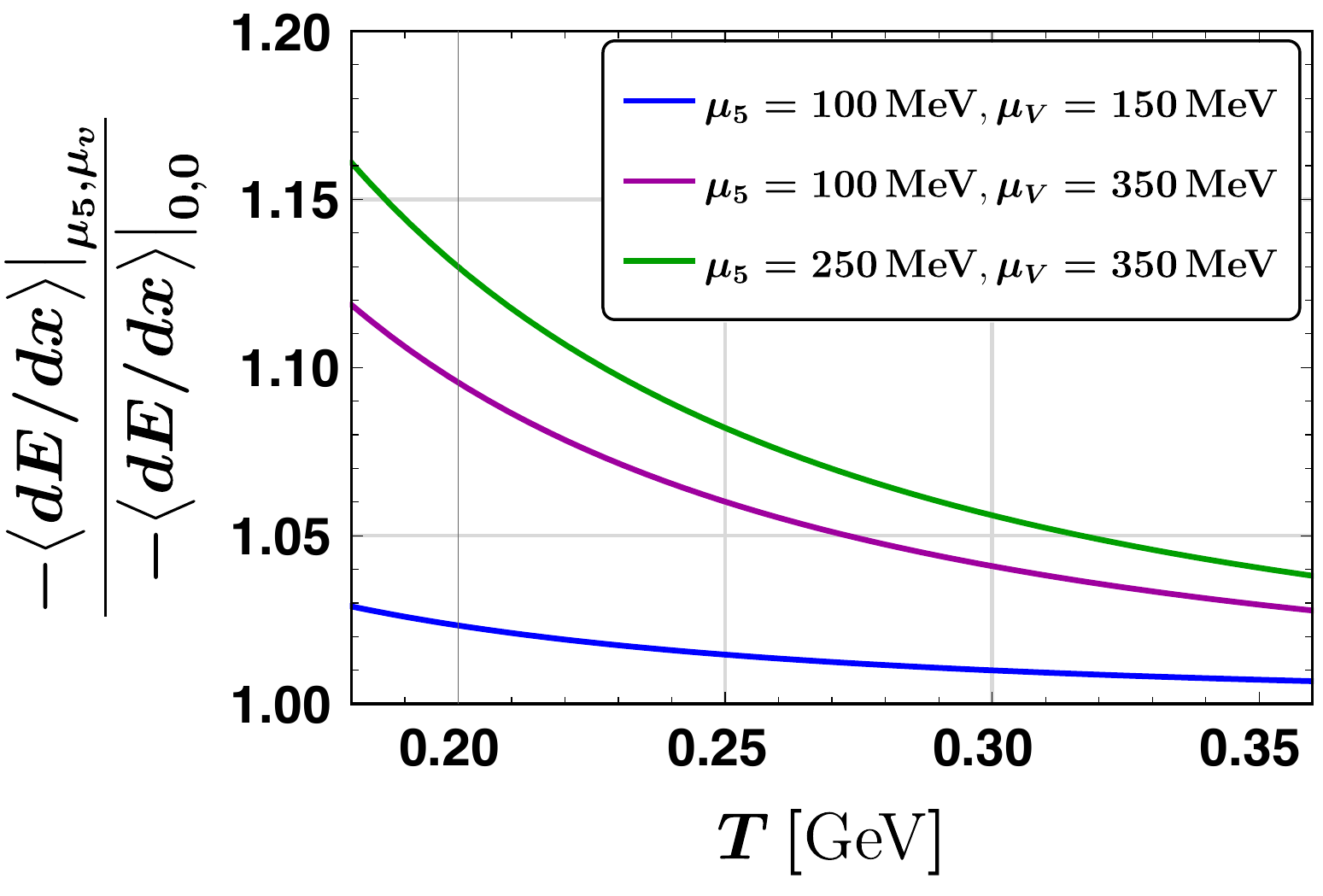}
			\caption{ Temperature dependence of the energy loss of a fast-moving parton in a chiral imbalance medium (left panel). The ratio of energy loss in a chiral medium to that of an isotropic medium is plotted as a function of temperature (right panel). }
			\label{fig:El_vs T}
		\end{center}
	\end{figure*}
In the presence of a magnetic field, defining $\Tilde{n}^i=A^{ij}n^j$ vector to describe the tensor basis is convenient. Here, $a$, $b$, $c$, and $d$ are Lorentz-invariant form factors and can be obtained from the following relations,
\begin{align}
&a = \frac{1}{2}N^{ij} \Pi^{ij},\\
&b = B^{ij} \Pi^{ij},\\
&c = R^{ij} \Pi^{ij},\\
&d = Q^{ij} \Pi^{ij}.
\end{align}
The form factors can be calculated from the one-loop gluon self-energy. The fermion loop is shown in Fig.~\ref{fig:se_B0}, which is affected by the magnetic field. We are considering the strong magnetic field limit with the energy hierarchy $T<\sqrt{|q_fB|}$. The form factors containing the gluon loop, ghost loop, and magnetic field modified quark loop contributions are given as~\cite{Karmakar:2018aig},
{\begin{align}
a =& \sum_f \frac{g_s^2 q_fB}{4\pi^2}e^{{-k_\perp^2}/{2q_fB}}~\frac{\om\,k_z}{\om^2-k_z^2} \sqrt{\frac{k_\perp^2 K^4}{\om^2\,|{\bf k}|^4}},\label{d1_sf}\\
b =&\frac{N_c\,g_s^2T^2}{3}\bigg(\frac{K^4}{\om^2\,|{\bf k}|^2}\bigg)\left[1-\mathcal{T}_K(\om,|{\bf k}|)\right]\nonumber\\
&-\sum_f \frac{g_s^2 q_fB}{4\pi^2}\bigg(\frac{K^4}{\om^2\,|{\bf k}|^2}\bigg)e^{{-k_\perp^2}/{2q_fB}}~\frac{k_z^2}{\om^2-k_z^2}, \label{b_sf} \\
c =&\frac{N_c\,g_s^2T^2}{3}\frac{1}{2}\left[\frac{\om^2}{|{\bf k}|^2}-\frac{K^2}{|{\bf k}|^2}\mathcal{T}_K(\om,|{\bf k}|)\right], \label{c_sf} \\
d =& \frac{N_c\,g_s^2T^2}{3}\frac{1}{2}\left[\frac{\om^2}{|{\bf k}|^2}-\frac{K^2}{|{\bf k}|^2}\mathcal{T}_K(\om,|{\bf k}|)\right]\nonumber\\
&+\sum_f \frac{g_s^2 q_fB}{4\pi^2} \bigg(\frac{K^4}{\om^2\,|{\bf k}|^2}\bigg)e^{{-k_\perp^2}/{2q_fB}}~ \frac{k_z^2}{\om^2-k_z^2}, \label{d_sf}
\end{align}}
where the angular integral reads as $\mathcal{T}_K(\om,|{\bf k}|)=\int \frac{d\Omega}{4\pi}\frac{\om}{\om-\bold k \cdot \hat p}$
with $\hat p$ is unit vector along $\bf{ p}$,  $i.e.$, $\hat p=\bf {p}/|\bf p|$ and $k_\perp=\sqrt{k_x^2+k_y^2}$. The general structure of the gluon effective propagator in the magnetized medium can be written as,
	\begin{align}
	   \Delta^{ij}= C_1\, B^{ij}+C_2\, R^{ij}+C_3\, Q^{ij}+C_4\, N^{ij},
	\label{eff_prop} 
	\end{align}	
where $C_{1,..4}$ are related to the form factors as,
\begin{align}
	   &C_1= \frac{K^2-d}{(K^2-b)(K^2-d)-a^2}, \nonumber\\
    &C_2= \frac{1}{K^2-c},\nonumber\\
	&C_3= \frac{K^2-b}{(K^2-b)(K^2-b)-a^2}, \nonumber\\
	&C_4=\frac{a}{(K^2-b)(K^2-d)-a^2}.
	\label{eff_prop1} 
	\end{align}	
Employing Eq.~\eqref{eff_prop} in Eq.~\eqref{main}, we can estimate the parton energy change in the magnetized medium as,
\begin{align}
    \bigg\langle{\frac{d{ E}}{dx}}\bigg\rangle=&\frac{ ig_s C_F}{{ |{\bf v}|}}
    \int \frac{d^3{\bf k}}{(2\pi)^3}{\omega }\Bigg\{C_1\frac{\omega^2}{|{\bf k}|^2}-C_2\bigg({ |{\bf v}|}^2-\frac{\omega ^2}{|{\bf k}|^2}\nonumber\\&+
    \frac{1}{\Tilde{n}^2}({\bf v}\cdot{\bf n})^2+2\frac{k_z\omega}{\Tilde{n}^2\, |{\bf k}|^2}({\bf v}\cdot{\bf n})+\frac{k_z^2\, \omega^2}{|{\bf k}|^4\Tilde{n}^2}\bigg)\nonumber\\&
    +\frac{C_3}{\Tilde{n}^2}\bigg(({\bf v}\cdot{\bf n})^2+2\frac{k_z\omega}{|{\bf k}|^2}({\bf v}\cdot{\bf n})+\frac{k_z^2\, \omega^2}{|{\bf k}|^4}\bigg)\nonumber\\&
    -\frac{2C_4}{\sqrt{\Tilde{n}^2}}\frac{\omega}{|{\bf k}|}\bigg(({\bf v}\cdot{\bf n})+\frac{k_z\, \omega}{ |{\bf k}|^2}\bigg)\Bigg\}.
    \label{mag2}
 \end{align}
Here, we define $({\bf k}\cdot{\bf n})=k_z$ as the magnetic field is fixed along the $\hat{z}-$direction in the present analysis. The term $({\bf v}\cdot{\bf n})$ describes the relative orientation of the parton with respect to the direction of the magnetic field and has a direct influence on the energy loss. This can give rise to anisotropy in the energy loss of the parton in the magnetized medium. In addition to that, the magnetic field effects are entering through the form factors as described in Eqs.~\eqref{d1_sf}-\eqref{d_sf}. For the perpendicular case, where the parton is moving transverse to the direction of the magnetic field, $i.e.$, $({\bf v}\cdot{\bf n})=0$, Eq.~\eqref{mag2} can be further simplified as, 
\begin{align}
    \bigg\langle{\frac{d{ E}}{dx}}\bigg\rangle=&\frac{ ig_s C_F}{{ |{\bf v}|}}
    \int \frac{d^3{\bf k}}{(2\pi)^3}{\omega }\Bigg\{C_1\frac{\omega^2}{|{\bf k}|^2}-C_2\Big({ |{\bf v}|}^2-\frac{\omega ^2}{|{\bf k}|^2}\Big)\nonumber\\&
    +(C_3-C_2)\frac{k_z^2\, \omega^2}{|{\bf k}|^4\Tilde{n}^2}
    -\frac{2C_4}{\sqrt{\Tilde{n}^2}}\frac{k_z\, \omega^2}{\, |{\bf k}|^3}\Bigg\}.
    \label{mag3}
 \end{align}
Next, we shall discuss the results obtained from various sections.

\section{Results}\label{Res}
Our primary findings are the energy loss of a test parton in the presence of chiral imbalance and a strong magnetic field in the QGP medium. The impact of chiral asymmetry and medium temperature over the energy loss of the fast-moving parton within the massless limit is depicted in Fig.~\ref{fig:El_vs T} (left panel). { In this figure, the imbalance between right-handed and left-handed fermions is measured using a chiral chemical potential $\mu_5=\mu_R-\mu_L$. For the quantitative estimation, we consider $\mu_L^2+\mu^2_R=\frac{1}{2}(\mu_5^2+\mu_V^2)$ with $\mu_V=\mu_R+\mu_L$.} The induced current due to the passage of the parton depends on the chirality of the medium and is reflected in the energy loss mechanism as described in Eq.~\eqref{Ch4}. The passage of the energetic parton is influenced by the chiral effects of the medium that enter through the medium's screening mass and parity-violating component of the self-energy $\Pi_A$ as defined in Eq.~\eqref{Ch6} and Eq.~\eqref{Ch5}, respectively. Similar to the case of an isotropic medium, the parton energy loss in the chiral imbalance medium increases with an increase in the temperature of the medium. We observe an increment in the energy loss in the presence of the chiral asymmetry compared to the isotropic QGP medium with vanishing chemical potential,, which has a strong dependence on the values of $\mu_R$ and $\mu_L$. The ratio of  energy loss of a massless parton in the chiral imbalance medium to the isotropic medium is shown in Fig.~\ref{fig:El_vs T} (right panel). It is observed that the chiral effects of the medium on the parton energy loss goes down with the increase in temperature, whereas the impact is more visible in the low-temperature regime. 

The energy loss of a parton in a left- and right-hand dominated medium at a fixed temperature is plotted in Fig.~\ref{fig:contour_mu5}. The region above and below the diagonal line $\mu_R=\mu_L$ represent the left- and right-hand fermions dominated regions, respectively. Notably, the energy loss is symmetric at a fixed temperature in both regions. This is because the parton energy loss in chiral medium remains as a parity conserving term as $-\big\langle{\frac{d{ E}}{dx}}\big\rangle$ depends on $\mu_5^2$ as described in Eq.~\eqref{Ch4}. It is seen that the parton energy loss increases with the chemical potential. The curved lines that represent constant energy loss contours indicate that $-\big\langle{\frac{d{ E}}{dx}}\big\rangle$ in a chiral medium is not only depending on the value of $\mu_R$ and $\mu_L$ but also the difference between them.

	\begin{figure}
		\begin{center}
			\includegraphics[scale=0.55]{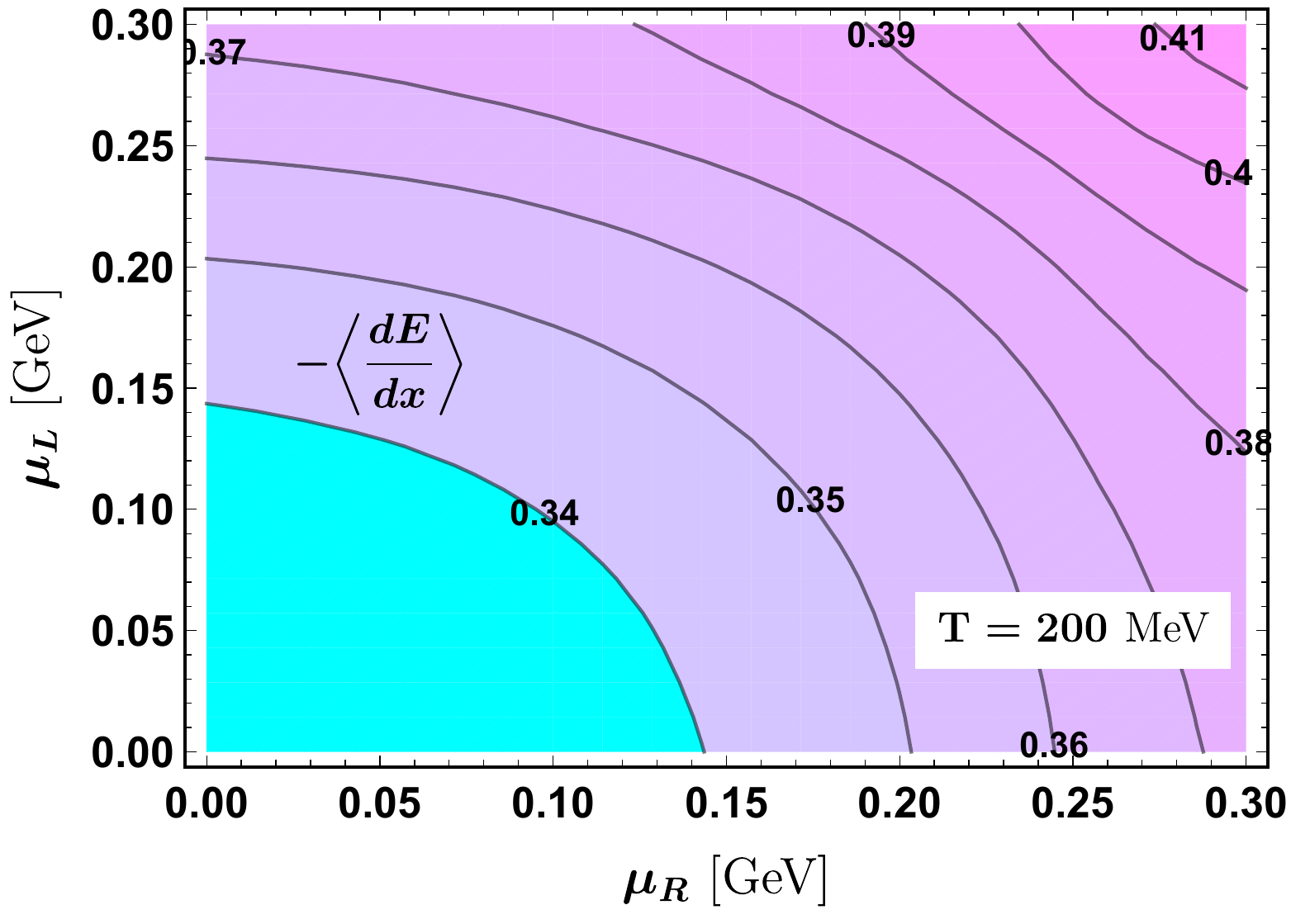}
			\caption{ Energy loss of a massless energetic parton in the chiral medium at $T=200$ MeV. Curved lines denote constant
energy loss contours and $\big\langle{\frac{d{ E}}{dx}}\big\rangle$ is described with the unit [GeV/fm].}
			\label{fig:contour_mu5}
		\end{center}
	\end{figure}
	\begin{figure}
			\includegraphics[scale=0.55]{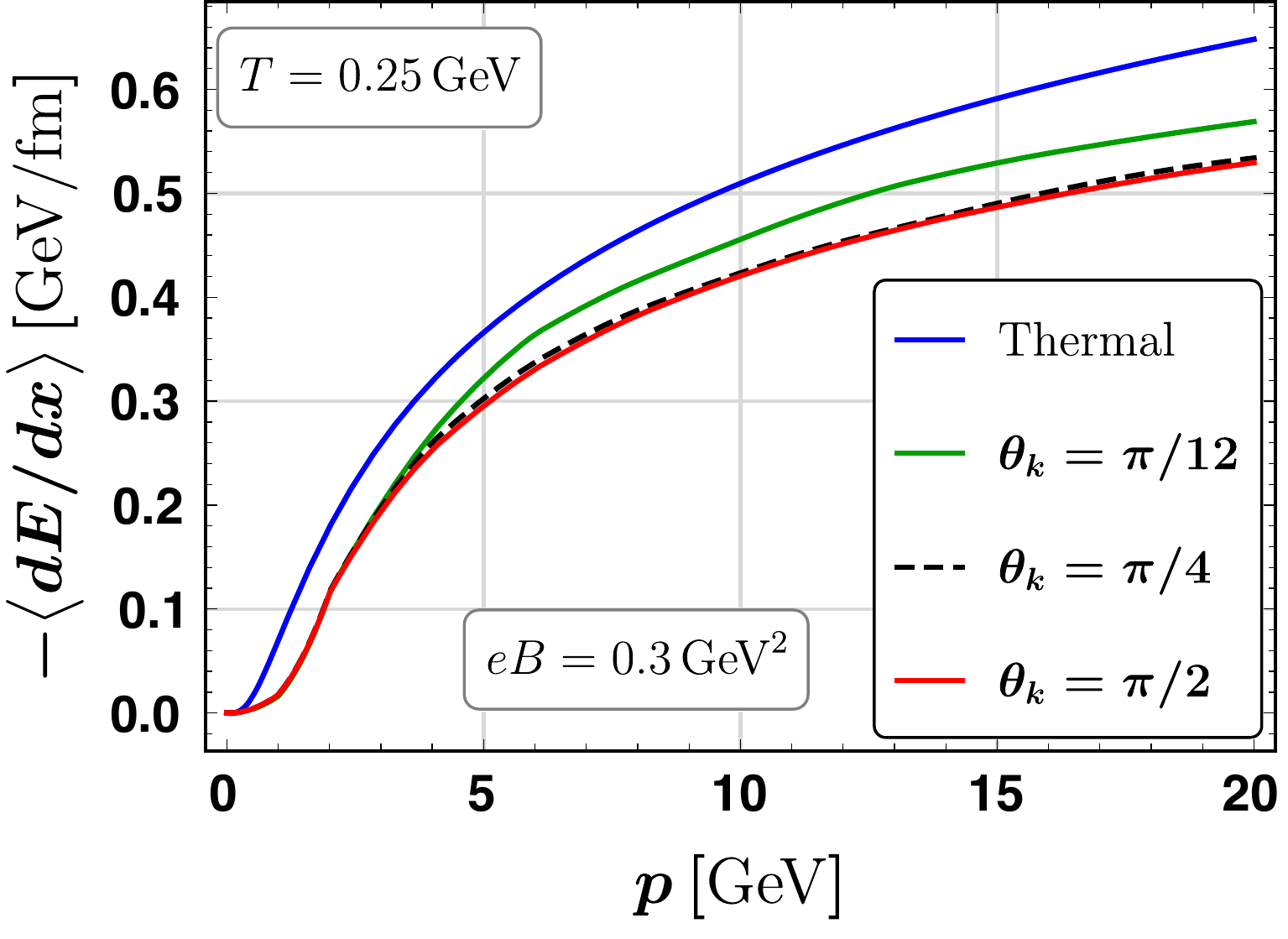}
			\caption{Momentum dependence of the parton energy loss in a magnetized QCD medium for various choices of angle between parton velocity and magnetic field in the medium ($\theta_k$). The mass of fast moving parton is chosen as $m=1.25$ GeV.}
			\label{fig:dedx_vs_p}
	\end{figure}

In Fig.~\ref{fig:dedx_vs_p}, we illustrated the momentum behavior of anisotropy generated due to the strong magnetic field to the energy loss mechanism. It is more relevant to the case of heavy quarks than the massless test parton, as they expect to witness a strong magnetic field in the initial stages of the collision. Here, we target the charm quark energy loss case with $m=1.25$ GeV, $T=250$ MeV, and $eB=0.3$ GeV$^2$ to ensure the strong field approximation. It is crucial to underline that the heavy quark does not undergo Landau-level dynamics owing to its large mass and initial momentum in the medium. However, the QCD medium properties and hence, the polarization effects will change with the inclusion of the magnetic field. They can indirectly affect charm quark energy loss in the magnetized medium. We observe that the energy loss of the charm quark traveling through a magnetized medium is sensitive to the magnetic field in the plasma. Notably, the energy loss strongly depends on the relative direction of the charm quark's velocity and magnetic field in the medium. The impact is more pronounced at the high momentum regime. It is seen that the energy loss is more when the motion is transverse to the direction of the magnetic field. 

The dependence of the strength of the magnetic field and temperature on the energy loss of a massless parton is shown in Fig.~\ref{fig:contour_eB}. We fix the relative orientation of parton motion and magnetic field by choosing $\theta_k=\pi/2$. { For an energetic parton, the magnetic field suppresses the energy loss compared to that in the thermal medium due to the dimensionally reduced Landau level motion of the charged fermions in the plasma.} The impact of the strength of the magnetic field is more pronounced in the low-temperature regimes. As the temperature rises, the suppression of parton energy loss that arises from the Landau kinematics is not sensitive to the strength of the magnetic field. The observation holds true for other choices of $\theta_k$ as well. 
	\begin{figure}
		\begin{center}
			\includegraphics[scale=0.55]{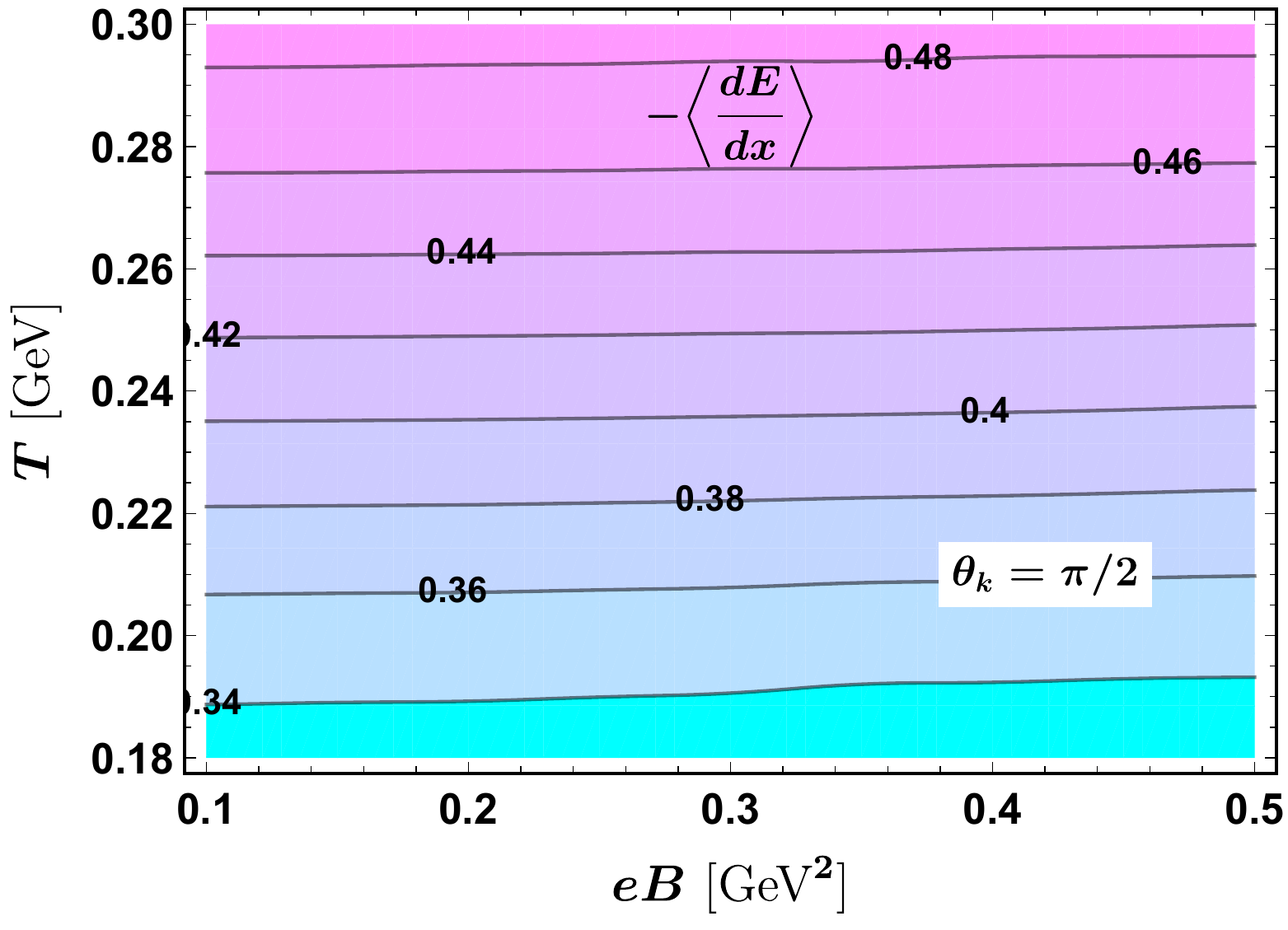}
			\caption{Energy loss (in [GeV/fm]) of an  energetic massless parton as a function of temperature and magnetic field.}
			\label{fig:contour_eB}
		\end{center}
	\end{figure}
\section{Summary and Outlook}\label{Summ}

In conclusion, we have studied the energy loss of a fast-moving parton for three scenarios within the QGP medium, {\it viz.,} the isotropic, chiral asymmetric, and strongly magnetized QGP medium. To do so, we employed Wong's equations combined with linearized Yang-Mills equations that describe the back-reaction exerted on the energetic parton by the medium while traversing through it. In the isotropic case, it is found that the energy loss increases with the temperature of the medium. It further increases with the momentum of the test parton. Similar trends are found for the chiral asymmetric medium as well.
Moreover,  the energy loss turned out to be sensitive to the chiral chemical potential, especially at the low-temperature regimes. We further investigated the heavy parton energy loss in a magnetized QCD medium. It is noticed that the magnetic field induces anisotropy in the medium that, in turn, suppresses the energy loss. Furthermore, the energy loss is found to have a strong dependence on the relative orientation of the parton's velocity and the magnetic field. It is seen parton loses more energy when it travels transversely to the direction of the magnetic field compared to other directions.

{ The current focus is on the parton energy loss due to the polarization effects of the medium. To estimate the complete energy loss of the test parton in the medium, this should be combined with the contribution} which originates from the hard elastic scattering of the parton with the constituents of the medium that may have a strong dependence on the magnetic field in the medium. Another interesting direction is to set up the formalism by considering the fluctuations in a magnetized QCD medium and exploring the possibility of energy gain of the parton. These are interesting aspects to explore shortly.

\section*{Acknowledgments}
We thank Mayank Singh for helpful discussions and feedback.
R.G. acknowledges financial support from the Department of Atomic Energy,
India. M.Y.J would like to acknowledge the SERB-NPDF
(National postdoctoral fellow) fellowship with File No.
PDF/2022/001551. M.K acknowledges the support from Special Postdoctoral Researchers Program of RIKEN.

\bibliography{main2}

\end{document}